\documentclass[twocolumn,showpacs,amsmath,amssymb]{revtex4}
\usepackage{epsfig}

\begin{document}

\title{Slow-light enhancement of Beer--Lambert--Bouguer absorption}

\author{Niels Asger Mortensen and Sanshui Xiao}

\affiliation{MIC -- Department of Micro and Nanotechnology,
Nano$\bullet$DTU, Technical University of Denmark, DTU-building
345 east, DK-2800 Kongens Lyngby, Denmark}

\date{\today}

\begin{abstract}
We theoretically show how slow light in an optofluidic environment
facilitates enhanced light-matter interactions, by orders of
magnitude. The proposed concept provides strong opportunities for
improving existing miniaturized chemical absorbance cells for
Beer--Lambert--Bouguer absorption measurements widely employed in
analytical chemistry.
\end{abstract}

 \maketitle

Optical techniques are finding widespread use in chemical and
bio-chemical analysis, and Beer--Lambert--Bouguer (BLB) absorption
in particular has become one of the classical work\-horses in
analytical chemistry~\cite{Skoog:1997}. During the past decade,
there has been an increasing emphasize on miniaturization of
chemical analysis systems~\cite{Janasek:2006} and naturally this
has stimulated a large effort in integrating
microfluidics~\cite{Squires:05,Whitesides:2006} and optics in
lab-on-a-chip microsystems~\cite{Verpoorte:2003}, partly defining
the emerging field of
optofluidics~\cite{Psaltis:2006,Monat:2007a}. At the same time,
there is an increasing attention to slow-light phenomena as well
as the fundamentals and applications of light-matter interactions
in electromagnetically strongly dispersive
environments~\cite{Lodahl:2004,Soljacic:2004,Vlasov:2005,Jacobsen:2006,Noda:2006}.
In this Letter we consider the classical problem of BLB
absorption. As with the phenomenon of photonic band-edge
lasing~\cite{Dowling:1994}, we show how slow light in an
optofluidic environment facilitates enhanced light-matter
interactions, by orders of magnitude, with strong opportunities
for improving existing miniaturized chemical absorbance cells.

\begin{figure}[b!]
\begin{center}
\epsfig{file=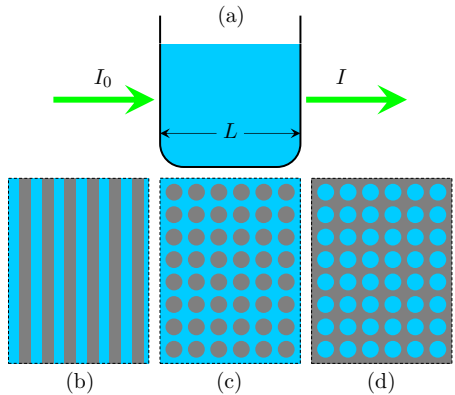, width=0.7\columnwidth,clip}
\end{center}
\caption{Schematic of (a) classical set up of
Beer--Lambert--Bouguer chemical absorbance cell and (b--d)
examples of strongly dispersive environments provided by photonic
crystals with dielectric regions (grey) with a dielectric function
different from that of the liquid sample (blue).} \label{fig1}
\end{figure}

The principle of a BLB measurement is illustrated in panel (a) of
Fig.~1 with an optical probe, with intensity $I_0$, incident on a
sample with absorption parameter $\alpha_l$ due to a concentration
of some chemical species. Typically the chemicals will be
dissolved in a liquid, but gas and transparent solid phases are in
principle also possible. Neglecting coupling issues, the
transmitted intensity $I$ will then, quite intuitively, be
exponentially damped, $I=I_0\exp(\gamma \alpha_l L)$, with $L$
being the optical path length and $\gamma$ being a dimensionless
measure of the slow-light enhanced light-matter interactions. For
a uniform medium of course, we have $\gamma\equiv 1$ and the
expression is often referred to as Beer's law. Since $\alpha$
correlates with the concentration of the absorbing chemical
species, Beer's law provides optical means for detecting and
quantifying the concentration of chemical
solutions~\cite{Skoog:1997}. Obviously, the effect relies heavily
on having a sufficiently long optical path length and the longer
$L$ is the lower a concentration can be monitored for a given
sensitivity of the optical equipment measuring $I/I_0$.
Lab-on-a-chip implementations of chemical absorbance cells are
thus facing almost exhausting challenges since the
miniaturization, i.e. reduction of $L$, decreases the sensitivity
significantly. This problem has already been confirmed
experimentally for lab-on-a-chip systems operated in the visible
with $L$ of the order $100$ to $1000\,{\rm \mu
m}$~\cite{Mogensen:2003}. In this work, we show a route to achieve
enhancement factors $\gamma$ much larger than unity, thus
potentially compensating for the cost of miniaturization and
reduction in optical path length.

\begin{figure}[t!]
\begin{center}
\epsfig{file=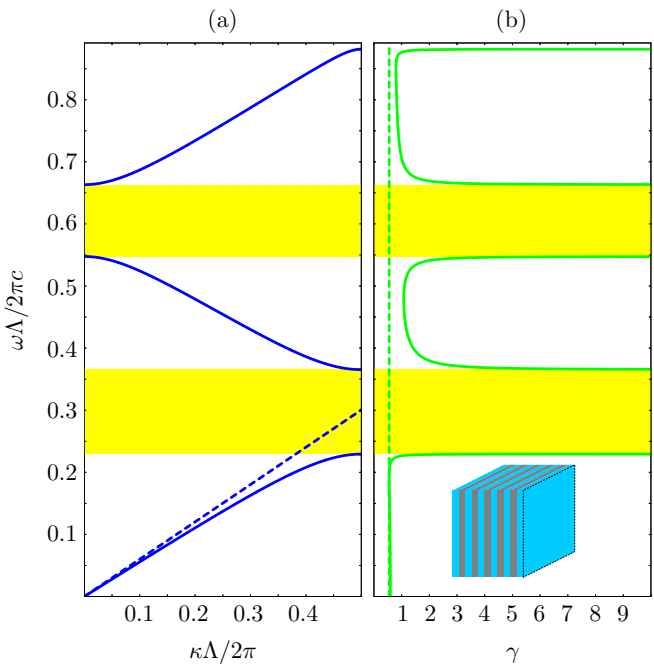, width=\columnwidth,clip}
\end{center}
\caption{(a) Photonic band structure for normal incidence of
either TE or TM polarized light on a Bragg stack of period
$\Lambda=a_l+a_2$ with $n_l=1.33$, $n_2= 3$, $a_l=0.8\Lambda$, and
$a_2 = 0.2\Lambda$. Photonic band gaps are indicated by yellow
shading and the dashed line indicates the long-wavelength
asymptotic limit where $\omega\simeq
c\kappa\Lambda/(a_ln_l+a_2n_2)$. (b) Corresponding enhancement
factor which peaks and exceeds unity close to the photonic
band-gap edges. The dashed line indicates the long-wavelength
asymptotic limit where $f\simeq a_ln_l^2/(a_l n_l^2+a_2 n_2^2)$.}
\label{fig2}
\end{figure}

In order to explicitly show this phenomenon, we start from the
electromagnetic wave equation for the electrical field,
\begin{equation}\label{eq:wave}
\nabla\times\nabla\times \big|E\big>=\epsilon
\frac{\omega^2}{c^2}\big|E\big>,
\end{equation}
and consider the case of a weakly absorbing medium with
$\epsilon=\varepsilon+i\delta\varepsilon$. Since absorption is a
weak perturbation, $\delta\varepsilon\ll\varepsilon$, standard
first-order electromagnetic perturbation theory is fully adequate
to predict the small imaginary shift in frequency,
$\Delta\omega\propto i\delta\varepsilon$. For the unperturbed
problem, we may obtain the dispersion relation $\omega(\kappa)$ by
solving the wave equation (see method section) with
$\epsilon=\varepsilon$. For a fixed frequency, the perturbation
$i\delta\varepsilon$ changes into an imaginary shift
$i\Delta\kappa$ of the wave vector $\kappa$ so that the absorption
parameter $\alpha=2\Delta\kappa$ becomes $\alpha=
k(c/v_g)\big<E\big|\delta\varepsilon\big|E\big>/\big<E\big|\varepsilon\big|E\big>$,
where the electrical field is the unperturbed field in the absence
of absorption, $v_g=\partial\omega/\partial\kappa$ is the group
velocity, and $k=\omega/c$ is the free-space wave vector. As a
reference we consider a homogeneous liquid with $\epsilon_l=n_l^2$
where we have a linear dispersion $\omega(\kappa)=(c/n_l)\kappa$
with a group velocity of $c/n_l$ and thus
$\alpha_l=k\delta\varepsilon/n_l$. Next, imagine that the
dispersion is modified by introducing a non-absorbing (at least
compared to the liquid) material of different index in the liquid,
see panels (b--d) in Fig.~1. Compared to the bare liquid such a
composite medium may support an enhancement of the effective
absorption. The enhancement factor $\gamma\equiv \alpha/\alpha_l$
can now be expressed as
\begin{equation}
\label{eq:gamma} \gamma =f\times\frac{c/n_l}{v_g},\quad f\equiv
\frac{\big<E\big|D\big>_l}{\big<E \big|D\big>}
\end{equation}
where we have introduced the displacement field
$\big|D\big>=\varepsilon \big|E\big>$. The integral in the
nominator of the filling factor $0<f<1$ is restricted to the
region containing the absorbing fluid while the integral in the
denominator is spatially unrestricted. This expression clearly
demonstrates how BLB absorption benefits from slow-light
phenomena. For liquid infiltrated photonic crystals and photonic
crystal waveguides, it is possible to achieve $v_g\ll c$ and at
the same time have a filling factor of the order unity, $f\sim 1$,
whereby significant enhancement factors become feasible. The
effective enhancement of the absorption can also be understood in
terms of an effective enhancement of the light-matter interaction
time given by the Wigner--Smith delay time $\tau$. For the
homogeneous problem, we have $\tau_l\sim L/(c/n_l)$ while for the
strongly dispersive problem $\tau\sim L/v_g$ so that
$\gamma\sim\tau/\tau_l \propto (c/n_l)/v_g$ in agreement with the
result in Eq.~(\ref{eq:gamma}) rigorously derived from
perturbation theory. The presence of the filling factor $f$ is
also easily understood since only the fraction $f$ of the light
residing in the fluid can be subject to absorption. These
conclusions may be extended to also non-periodic systems,
including enhanced absorption in disordered systems as well as
intra-cavity absorbance configurations, by use of scattering
matrix arguments~\cite{Beenakker:2001}.

\begin{figure}[b!]
\begin{center}
\epsfig{file=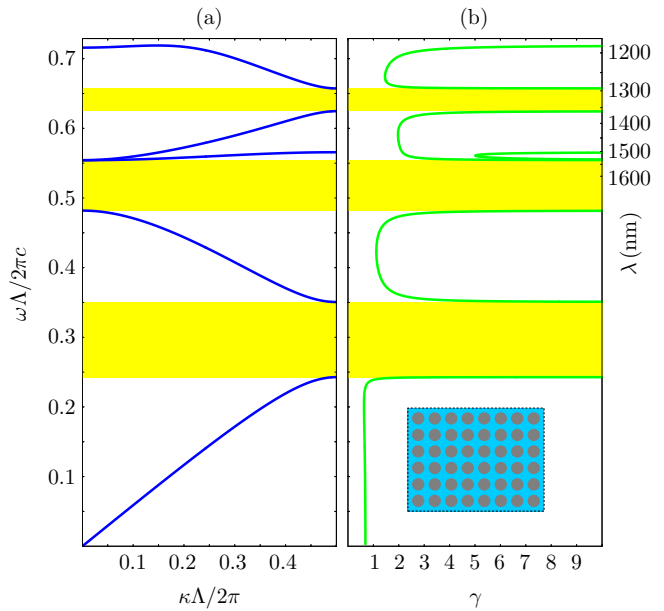, width=\columnwidth,clip}
\end{center}
\caption{(a) Photonic band structure for propagation of TM
polarized light along the $\Gamma$X direction in a square lattice
of period $\Lambda$ with dielectric rods of diameter
$d=0.4\Lambda$ and $\varepsilon=10.5$. Photonic band gaps are
indicated by yellow shading. (b) Corresponding enhancement factor
which exceeds unity for the flat bands in general and the third
band in particular. The right $y$-axis shows the results in terms
of the free-space wavelength when results are scaled to a
structure with $\Lambda=850\,{\rm nm}$. } \label{fig3}
\end{figure}

Let us next illustrate the slow-light enhancement for the simplest
possible structure; a Bragg stack with normal incidence of
electromagnetic radiation. Panel (a) of Fig.~2 shows the photonic
band structure of an optofluidic Bragg stack of period
$\Lambda=a_l+a_2$ with the low-index material layers of width
$a_l=0.8\Lambda$ being a liquid with refractive index $n_l=1.33$
while the high-index layers have a width $a_2 = 0.2\Lambda$ and a
refractive index $n_2= 3$. Photonic band gaps are indicated by
yellow shading and the dashed line indicates the long-wavelength
asymptotic limit where the Bragg stack has a meta-material
response with a close-to-linear dispersion $\omega\simeq
c\kappa\Lambda/(a_ln_l+a_2n_2)$. When approaching the band-gap
edges, the dispersion flattens corresponding to a slow group
velocity. It is well-known that the flat dispersion originates
from a spatial localization of the field onto the high-index
layers and thus $f\ll 1$ near the band edges where the inverse
group velocity diverges. However, in spite of the localization,
the enhancement factor may still exceed unity as shown in panel
(b) where the dashed line indicates the long-wavelength asymptotic
limit with $f\simeq
\big<1\big|\epsilon\big|1\big>_l/\big<1\big|\epsilon\big|1\big>=a_ln_l^2/(a_l
n_l^2+a_2 n_2^2)$. In order to further benefit from the slow-light
enhanced light-matter interaction, we obviously have to pursue
optofluidic structures supporting both low group velocity and at
the same time large filling factors. Fig.~3 shows one such example
where high-index dielectric rods are arranged in a square lattice.
Compared to the Bragg stack, some of the modes in this structure
have both a low group velocity and at the same time a reasonable
value of the filling factor $f$. Particularly the third band in
panel (a) is quite flat and with a finite $f$ giving rise to an
enhancement factor $\gamma$ exceeding 5 even at the centre of the
band. As indicated on the right $y$-axis, the enhancement may have
a bandwidth of order 50~nm for a pitch around $\Lambda\sim
850\,{\rm nm}$, which indeed makes fabrication of such structures
realistic with state of the art micro and nano fabrication
facilities. As a final example, Fig.~4 shows the result of
introducing a line-defect waveguide in such a structure. The
waveguide mode has $f\sim 60\%$ combined with a low group velocity
near the band edges.

\begin{figure}[t!]
\begin{center}
\epsfig{file=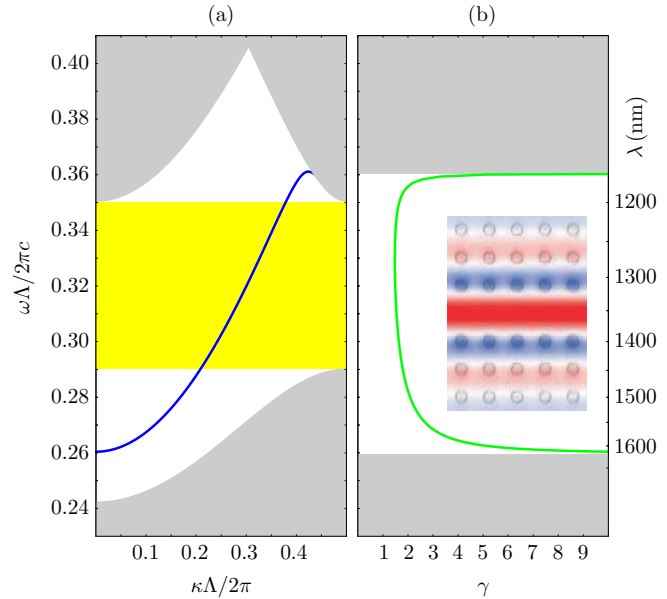, width=\columnwidth,clip}
\end{center}
\caption{(a) Photonic band structure for propagation of TM
polarized light along the $\Gamma$X direction in a line-defect
waveguide in a square lattice of period $\Lambda$ with dielectric
rods of diameter $d=0.4\Lambda$ and $\varepsilon=10.5$. The
complete photonic band gap of the photonic crystal is indicated by
yellow shading while grey shading indicates the finite
density-of-states in the photonic crystal due to the projected
bands in the Brillouin zone. (b) Corresponding enhancement factor
which exceeds unity over the entire bandwidth. The right $y$-axis
shows the results in terms of the free-space wavelength when
results are scaled to a structure with $\Lambda=420\,{\rm nm}$.
The inset shows the electrical field of the waveguide mode at the
$\Gamma$-point $\kappa=0$. } \label{fig4}
\end{figure}

For the above numerical results, fully-vectorial eigenmodes of
Maxwell's equations, Eq,~(\ref{eq:wave}), with periodic boundary
conditions were computed by preconditioned conjugate-gradient
minimization of the block Rayleigh quotient in a planewave basis,
using a freely available software package~\cite{Johnson:2001}. For
the resolution, we have used a basis of $2^{10}$ plane waves for
the 1D problem in Fig.~2 and $2^{10}\times 2^{10}$ plane waves for
the 2D problem in Fig.~3. In Fig.~4 we have used a resolution of
$2^{7}\times 2^{7}$ and a super-cell of size $1\times 7$.

In the above examples we have for simplicity considered dielectric
constants corresponding to semiconductor materials suitable for
the near-infrared regime. However, we would like to emphasize that
applications exist also in the visible, mid-infrared,
far-infrared, and even the microwave and sub-terahertz regimes.
The predicted enhancement of light-matter interactions makes
liquid-infiltrated photonic crystals obvious candidates for
improving existing miniaturized chemical absorbance cells.
Previous work on liquid-infiltrated photonic
crystals~\cite{Loncar:2003,Chow:2004,Kurt:2005,Erickson:2006,Hasek:2006}
has focused on the solid type with liquid infiltrated voids
illustrated in panel (d) of Fig.~1, while we in this work have
focused on rod-type photonic crystals which have the technological
strong advantage that they are permeable to an in-plane liquid
flow, thus making them integrable with micro-fluidic channels in
planer lab-on-a-chip technology.

In conclusion, we have studied the potential of using
liquid-infiltrated photonic crystals to enhance
Beer--Lambert--Bouguer absorption. The slow-light enhancement of
the absorption, by possibly orders of magnitude, may be traded for
yet smaller miniaturized systems or for increased sensitivity of
existing devices.

\emph{Acknowledgments.} This work is financially supported by the
\emph{Danish Council for Strategic Research} through the
\emph{Strategic Program for Young Researchers} (grant no:
2117-05-0037).

%\bibliographystyle{prsty}
%\bibliography{Q:/papers/BibTeX/OFTS}

\newpage

\end{document}